# Archiving: The Overlooked Spreadsheet Risk

Dr. Victoria Lemieux
IT Risk & Security
Credit Suisse First Boston
One Cabot Square, London, E14 4QJ
*Vicki.lemieux@csfb.com*

## ABSTRACT

*This paper maintains that archiving has been overlooked as a key spreadsheet internal control. The case of failed Jamaican commercial banks demonstrates how poor archiving can lead to weaknesses in spreadsheet control that contribute to operational risk. In addition, the Sarbanes-0xley Act contains a number of provisions that require tighter control over the archiving of spreadsheets. To mitigate operational risks and achieve compliance with the records-related provisions of Sarbanes-Oxley, the author argues that organisations should introduce records management programmes that provide control over the archiving of spreadsheets. At a minimum, spreadsheet archiving controls should identify and ensure compliance with retention requirements, support document production in the event of regulatory inquiries or litigation, and prevent unauthorised destruction of records.*

## 1. INTRODUCTION

Many companies rely on spreadsheets for financial reporting and support of operating processes. For example, in a financial services firm, spreadsheets may be used to perform reconciliations by downloading information from two systems into separate existing MS Excel spreadsheets. MS Excel functions and pivot tables are then used to create summary data for each source. When spreadsheets are core to business processes poor control over them can have a significantly negative effect upon companies' bottom line and reputation. PriceWaterhouseCoopers reports the following examples of how spreadsheet risks can impact upon the corporate bottom line (PwC, 2005):

-A spreadsheet error at a major financial institution was deemed a significant factor in a major $1billion financial statement error in the classification of securities. The error resulted from a flawed change control process - an unapproved change to the formula within the spreadsheet -and other control deficiencies, including lack of technical and user documentation, insufficient testing, and inadequate backup and recovery procedures.

-A utilities company took a $24million dollar charge to earnings after a spreadsheet error -simple mistake in cutting and pasting - resulted in an erroneous bid in the purchase of hedging contracts at a higher price than it wanted to pay.

-A trader at a bank was able to perpetrate fraud by manipulating spreadsheet models used by the bank's risk control staff. Because of inadequate controls over spreadsheet this fraud continued for months.

Not surprisingly then, being able to demonstrate sound internal control of critical spreadsheets in compliance with the Sarbanes-Oxley Act (SOX) is as important as being able to demonstrate it in respect of core processing applications and other critical systems. With the introduction of SOX, applying sound internal controls to spreadsheets not only makes good business sense but also becomes a legal requirement.

This paper will argue that archiving has been overlooked as a key spreadsheet internal control. Using a case study of failed Jamaican commercial banks, it will demonstrate how poor archiving can lead to weaknesses in spreadsheet control that contribute to operational risk. This will be followed by a discussion of SOX archiving requirements and how to mitigate archiving risks and introduce into an organisation best spreadsheet archiving practices for SOX compliance.

## 2. ARCHIVING: THE OVERLOOKED SPREADSHEET RISK

Discussions of spreadsheet-related risks generally focus on:

- Complexity of the spreadsheet and calculations
- Purpose and use of the spreadsheet
- Number of spreadsheet users
- Type of potential input, logic and interface errors
- Size of the spreadsheet
- Degree of understanding and documentation of spreadsheet requirements by the Developer
- Uses of the spreadsheet's output
- Frequency and extend of changes and modifications to the spreadsheet
- Development and developer training and testing before the use of the spreadsheet (PWC 2005).

Equally important, however, are the risks associated with failing to properly archive spreadsheets.

Why should spreadsheet archiving be considered a critical risk area? Simple: there are risks to the business when critical information is not properly retained and accessible, especially in the post-SOX world.

## 3. A CASE STUDY OF ARCHIVING RISKS

A study of the Jamaican banking crisis, in which all of the country's indigenous commercial banks failed, shows how poor control of spreadsheet archiving contributed to the flawed decision making that, in turn, fed into to a failure of Jamaican banks (Lemieux, 2002). Like many other firms, the banks relied heavily on spreadsheets because their major transaction processing and risk management systems failed to meet management information and reporting requirements.

Interviews with former employees of the failed banks reveal that they used spreadsheets in the following ways:

- Cash management
- Financial control and budgeting

- Analysis of customer and product profitability
- Analysis of the cost of funds
- Currency position management
- Credit decision management
- Interest rate sensitivity analysis and risk management
- Recording of proprietary trading in securities.

There was an ad hoc approach to spreadsheet archiving at these banks, characterised by:

- Individualistic naming of files. Individuals were allowed to assign their own names to files. These names often gave no clue as to the content of the file or its relation to a business process. Ultimately, individualistic naming of files was a major factor in the inability to locate important spreadsheets. Once the creator of the spreadsheet left a bank, the spreadsheet was as good as gone with them since the knowledge of its existence and how to retrieve it vanished with the individual who named and stored it. Even when the creator was still around, individually named spreadsheets became "information islands", often only available to the single user responsible for their creation even though the information they contained was of benefit to the decision-making processes of others.

- Ad hoc assignment of storage location. Individuals were permitted to store spreadsheets in their personal drives to which they alone had access and for which they alone made decisions about retention or deletion of documents.

- Absence of any objective criteria governing deletion of spreadsheets from storage. Without any clear understanding of the importance of spreadsheets in the management information and reporting process, no one saw any reason to control their deletion. This resulted in periodic purges of important spreadsheets as storage locations became full. Since there was no understanding of these spreadsheets as "records" that needed to be kept as evidence of how the bank's financial positions were calculated, when drives filled up and the notice came round from IT to make more space available, individuals purged their drives and wiped out the evidentiary trail.

- Failure to preserve a link to the business context in which the spreadsheets were created. This failure often rendered spreadsheets meaningless as background to a particular business decision.

- Inability to guarantee the authenticity and reliability of spreadsheets. Since there were no controls over how spreadsheets were archived and no effort was made to "lock" down their content as part of a formal archiving process, if anyone was lucky enough to actually locate one of these documents after a period of time, their integrity was seriously questionable, since anyone could have changed the content in the intervening period and audit trail controls were weak to non-existent.

Former managers at the failed banks commented on the impact of poor spreadsheet controls on the banks. One interviewee, for example, described how the bank was forced to rely on competitors' actions to drive its asset and liability management (ALM) policies because the banks own interest rate sensitivity data, which had been recorded in spreadsheets, could no longer be accessed. (Lemieux, 2002, p.258).

Poor control over spreadsheets at Jamaican indigenous banks contributed to management information and external reporting problems (i.e., P&L distortions) that contributed to the banks' management and external regulators losing sight of the banks' true positions and exposures. This problem fed a downward spiral into liquidity crisis.

As the Jamaican financial crisis unfolded, the government also recognised that fraud and corruption had contributed to the collapse of indigenous banks. To address these allegations, it established a team of foreign and local forensic auditors to work with the police fraud squad to identify and take action on instances of fraud. Inaccessibility of source documents, however, seriously hampered the auditors' work. The work of reconstructing what in some cases were very convoluted financial transactions was made extremely difficult by the fact that critical records, many in spreadsheet form, were missing. One interviewee said: "I am looking at a particular company now where I thought that I was told that all the servers there that I could run off the information. When somebody attempted to do that they realised that the diskette was bad or contaminated. So you have a whole year's [data] that you cannot access . . . I have tried so all I have now left to do is to utilise some of the hard copies. But it is not consistent. You have one month, you can't find two months and you have another month. So it is going to be very difficult to trace these transactions." (Lemieux, 2001, p.338).

Though the absence of archival controls over spreadsheets at the failed indigenous banks may have been extreme, similar problems are not unknown in other firms. Though no separate data exist for spreadsheets, the aggregated data covering all electronic records indicates that very few organisations have established formal programmes to systematically manage the archiving of their electronic records (A11M, 2005). Many still rely on back up processes better suited to disaster recovery than to the preservation of evidence to meet legal and regulatory requirements.

## 4. WHAT SOX SAYS ABOUT ARCHIVING

In the post-SOX world control over the archiving of spreadsheets becomes a critical compliance matter as well as a business competitiveness issue. There are a number of SOX provisions that impact upon records and information management, some more relevant to the matter of archiving spreadsheets than others. This section will highlight just a few of these (a full list of the requirements will be available in the appendix to the published version of this paper):

1. SOX 103(aX2)(A)(i) - Audit reports, work papers, and other information related to any audit report must be kept for at least 7 years; Audit reports must contain statements about the testing of internal controls and whether those control structures include maintenance of accounting records.
2. SOX 104(e) - Public accounting firms may be required to retain records not otherwise required under section 103.
3. SOX s. 802, rule 2-06(a) requires a 7 years after conclusion of the audit/review retention period for accountants to retain "records relevant to the audit or review of issuers' and registered investment companies' financial statements, including work papers and other documents that form the basis of the audit or review, and memoranda, correspondence, communications, other documents and records (including electronic records), which are created, sent or received in connection with the audit or review." A company, in consultation with its legal and accounting advisors, will need to determine which of its spreadsheets and other records fall within the meaning of this provision of the act and related rules. There is a 10-year penalty for violating this rule.

Given the records and information management requirements under SOX, express or implied, and the penalties for non-compliance, poor control over spreadsheet archiving is a risk that should not be left unmitigated.

## 5. ADDRESSING SPREADSHEET ARCHIVING RISK - ELEMENTS OF GOOD PRACTICE

It is best to address spreadsheet archiving as part of setting up (if there is no programme in place), maintaining, and ensuring compliance with an organisation-wide records management programme. This approach also will demonstrate that archiving controls are part of business as usual practice, not merely a "tick the box" approach to SOX compliance. It also will ensure that SOX controls do not work at cross-purposes with other organisational records requirements and controls.

SOX does not explicitly direct any records management activities, so organisations cannot simply follow a statutory recipe to achieve good archiving practices. However, the Act demands a number of specific outcomes that need to be underpinned by effective records and information management practices. The requirements of an organisation's records management programme should be guided by what is needed to meet these outcomes. International standards, such as ISO 15489, the International Records Management Standard, can provide guidance on how to go about establishing a compliant records management programme (ISO, 200 1). At the end of the day, the goal of the records management programme should be to create the processes, procedures and records necessary to demonstrate compliance with SOX and to repudiate any claims of misfeasance or malfeasance (Montana et al., 2003).

## 6. RETAINING RECORDS

Good records management practice calls for the establishment of Records Retention Schedules. These are documents that identify the records that must be created by law or regulation, and the period of time for which those records must be retained. An organisation should definitely have one of these documents in place.

Often, one finds that an organisation's archiving function has established a Records Retention Schedule, but that the scope of its coverage only extends to paper documents. There is no question that SOX requirements apply not just to paper records, but also to documents in a multitude of electronic forms, including electronic versions of spreadsheets. Indeed, ever since electronic forms of documents have become ubiquitous, the U.S. courts have shown a distinct favouritism for the submission of evidence in its "native" form (i.e., electronic) rather than receiving a paper "copy" (Wallace, 2001). Consequently organisations should be clear that their Records Retention Schedules apply to records in all forms.

In terms of the retention requirements related to SOX, there seems to be much debate and confusion in this area. Recent discussions on the IT Governance listserv led to a wild claim that all records had to be retained for 7 years. This is not the case. Sections 103(a)(2)(A)(i) and 802(1)(a) apply to audit records and audit work papers of public accounting firms. Industry best practice has evolved to include internal audit records and work papers as well, though there is no explicit requirement for retention of these records in the Act or its related regulations. Initially, the Act required that audit work papers be kept for a period of 7 years under section 103 and 5 years under section 802. The requirements under section 802 have subsequently been raised via U.S. Securities and Exchange Commission regulation to 7 years in order to harmonise with section 103 of the Act and with auditing standards (SEC, 2003). However, the Act also says that (s.802(2)(c)) nothing in it should be taken to diminish or relieve an obligation to comply with the records retention requirements or prohibitions on document destruction mandated by other legislation. This means that if audit records fall within the retention requirements of other legislation and those retention requirements are longer, the longer of the requirements would apply.

Audit-related records are the beginning and end of explicit retention requirements in SOX, but clearly the letter of the Act requires compliance with the retention requirements of other legislation (s. 802(2)(c)above). Moreover, the requirements of section 404 are underpinned by evidence of the establishment and proper operation of effective internal controls. This implies a much wider obligation on organisations to retain records. Many of these records will be spreadsheets created as part of SOX-relevant business processes. Though the Act focuses on the accuracy of corporate financial records, an organisation would be foolish to stop its records management efforts with financial records. The fact is that non-financial records can provide evidence of financial vulnerabilities. As such they will be deemed relevant to any SOX-related requirements and inquiries.

Another SOX-records retention myth that needs to be explored and, in my view, exploded, is that SOX-relevant records must be gathered up and kept in a single, SOX records repository. This would certainly be one approach to ensuring retention of the records required by SOX. The expense, however, could be prohibitive. Aside from the expense of gathering up all SOX-relevant records for retention, there is the question of whether removing the records from their business context has the potential to diminish the evidentiary qualities of the records. Unless carefully procedurally controlled, there easily could be a danger of reduced record integrity. Therefore, a better approach is to identify the records, properly manage and archive them "in situ" (i.e., within a production environment) or a corporate archiving environment, and apply appropriate indexing for retrieval.

Having said that Records Retention Schedules apply to records in all forms, including spreadsheets, the trick is putting them into effect. Most organisations retain electronic spreadsheets, as in the Jamaican case study above, on a variety of servers, and leave control over the life cycle of such documents to the individuals who generated them in the first place (i.e., usually an end user). In the post-SOX environment, it should be abundantly clear that this approach is no longer advisable.

Steps must be taken to ensure that spreadsheet content, structure and context, that is the links to the business transactions that they were created to support, are retained for their required period of time in a form acceptable to regulators, investigators and the courts. This paper will return in a later section to the implementation of Records Retention Schedule; but for now, suffice it to say, regulators have shown a definite impatience with organisations that are not able to produce requested documentation. For example, in 2002 the U.S. Securities and Exchange Commission levied fines against five investment banks for failure to preserve emails (SEC, 2002).

Records Retention Schedules must not only be implemented, they must be regularly reviewed. A regular review will ensure that the Records Retention Schedule remains consistent with legal and regulatory requirements, complete (i.e., incorporates the records generated from new business functions), and appropriate to the business environment. When reviewing Records Retention Schedules, therefore, organisations should look at the currency of its retention periods, records series, nomenclature, indexing and structure and overall compliance. A regularly reviewed Records

Retention Schedule will be a good defence strategy in the event of the kind of scrutiny that records management programmes can now come under in the event of SOX-related investigation or litigation.

## 7. RETRIEVING RECORDS

Documentation must be capable of being accurately and quickly retrieved in the event of an investigation, regulatory inquiry or litigation. Even the most complete documentation loses value if it cannot be retrieved. Given the climate of suspicion that was the impetus of the Act, delays in the production of legitimately requested records and information can be extremely damaging, Opponents and, it should be added, the public, are willing to assume bad motive and push for sanctions. For example, recently, a Florida court penalised the investment bank Morgan Stanley for "bad faith" actions in respect to handing over backup tapes containing emails relevant to the Perelman litigation against Sunbeam. The judge told the jury it should simply assume that Morgan Stanley helped **defraud** Mr. Perelman (Craig, 2005). Even without regulatory sanction or court-imposed penalties, the reputational damage can be significant.

The ability to accurately and quickly retrieve documentation will be assisted by setting up standardized file structures, implementing file naming conventions, and indexing spreadsheets. Personnel must also be sufficiently well trained to carry out document requests efficiently and in the time frame demanded, as well as to understand the need to protect the integrity of the documents throughout the retrieval process.

File naming conventions deserve special attention because one of the most common failings of retrieval systems is due to poor nomenclature. Under the best of circumstances, poor nomenclature impedes the ability of users to retrieve information efficiently. Under the worse case scenario, it can be interpreted more sinisterly. Poor naming conventions may be taken as an attempt to conceal information or, as in the recent Citigroup European Bond Trading Scandal wherein the bank's highly contentious bond trading move was rather ominously and unfortunately named "Dr. Evil" (Wall Street Journal, 2005), arm opponents and cause damage to a Firm's reputation. Clarity and transparency should be key goals in the development of file naming conventions and indexing plans.

Many organisations fail to preserve the links between individual documents and the business context to which they relate. This can be a mistake as it can render a document difficult to locate, open the meaning of the document up to "creative" interpretation by adversaries, and render it difficult to determine whether a document legitimately falls within the scope of a document production order or legal discovery exercise. Classification of documents according to corporate records taxonomy can serve as a vehicle for preserving contextual links in documents. Another means of achieving this goal is to capture contextual metadata, such as the name of the business process or transaction for which the document is being created, which is either stored in a database or embedded in the document itself. Business process flows can be another very useful way of capturing information about how spreadsheets fit into the overall business context.

## 8. DESTROYING RECORDS

One might think, given the harsh penalties associated with records destruction, that, record management programme or no records management programme, any records destruction should be halted. Quite to the contrary, records destruction should still take place in keeping with best practice, but it must be as part of the normal and ordinary course of business. The best way to demonstrate that legitimate destruction is for records disposals to take place in the context of an established records management programme and with full audit trails of disposal actions. 1n contrast, for any organisation undergoing Sarbanes-Oxley scrutiny, ad hoc destruction of records in the absence of a formal programme, no matter how innocent the motive, invites the most damning inferences as to reasons." (Montana et al., 2003).

ISO 15489, the International Records Management Standard, establishes the following principles governing records disposal (ISO, 200l):

- Disposition authorities that govern the removal of records from operational systems should be applied to records on a systematic and routine basis in the course of normal business activity.

- No disposition action should take place without the assurance that the record is no longer required, that no work is outstanding and that no litigation or investigation is current or pending (or even reasonably foreseeable) which would involve relying on the records as evidence.

- Destruction should always be authorised.

- Records pertaining to pending or actual litigation should not be destroyed.

- Records destruction should be carried out in a way that preserves the confidentiality of any information they contain.

- All copies of records that are authorised for destruction, including security copies, preservation copies and backup copies, should be destroyed.

- Records systems should be capable of facilitating and implementing decisions on retention or disposition of records

- It should be possible for these decisions to be made at any time in the existence of the records including during the design stage of the records systems.

- It should also be possible, where appropriate, for disposition to be activated automatically.

- Systems should provide audit trails or other methods to track completed disposition actions.

Given the penalties in the Act associated with destruction of records, it is worth focusing some detailed attention on the subject of destruction bans and legal holds on records destruction. The Act has established a Public Company Accounting Oversight Board (PCAOB). The PCAOB is vested with broad power to oversee public accountancy, set standards for the conduct of audits and maintenance of records by public accountants, and generally to oversee and enforce standards of public accounting.

The parties most directly affected by the PCA013 are public accountants and auditors, but the PCA013 also has investigative authority over the auditing of public companies. In general, the PCAOB is empowered to request and/or subpoena documents in the possession of any person, including a client of a registered public accounting firm, which the Board considers relevant or material to an investigation. A publicly traded company, thus, may find itself required to respond to an investigation by the PCAOB by the production of documents and information related to an audit. It is therefore imperative for organisations to have in place policies, procedures and systems for handling information production demands arising out of government investigations, litigation, and other legal and adversarial situations. Even for organisations that already have such policies, procedures and systems in place, the Act stipulates requirements that should encourage a check for efficacy in relation to the following (Montana, et al, p. 18):

- The Act grants authority to demand production of testimony or documents well in advance of any formal proceeding such as litigation. Are procedures sufficient to ensure that documents are safeguarded from the moment such proceedings are reasonably foreseeable?

- Are existing procedures sound and foolproof? For example, once a document destruction hold order has been issued, is the organisation confident that no documents will be destroyed. In many circumstances, implementation of a document destruction hold is the responsibility of the employee and there are very few controls in place to ensure this responsibility will be carried out.

- Is staff training and awareness sufficient to provide documents in the timeframe and with the accuracy required?

## 9. ARCHIVING STRATEGIES: SOME PROPOSALS

Having discussed general good practice in respect to archiving, this paper will now turn to discussing the specifics of how these practices might be applied to spreadsheet archiving.

SOX essentially requires that spreadsheets be dealt with as any other record that would be required as evidence to substantiate an organisation's financial statements. All records have a life cycle consisting of the following phases:

1. Creation and/or receipt
2. Active use
3. Semi-active use, during which records are referred less often because the business transaction for which they were created or received has been completed.
4. Inactive use, during which records are rarely referred to but must be retained for legal/regulatory or business reasons. It is during the inactive period that records generally are migrated from production environments to an archive and/or deleted from production environments.

The records life cycle roughly parallels the software life cycle. As with other types of records, effective spreadsheet archiving will begin at the point of spreadsheet creation and end only when the spreadsheet has met all retention requirements.

Like Word documents or MS Access databases, spreadsheets are created using end-user processing technology readily available on the desktop. As such, spreadsheets are often created and managed by the end-user, who may be very unfamiliar with the principles of managing the software or records life cycle. In some cases, however, because the spreadsheet performs quite complex processing functions and forms a critical bridge between applications in key business processes, an organisation's IT department may become involved in the design of the spreadsheet or aspects of its management. For this reason, it can be helpful to classify spreadsheets into two broad categories as follows so that the archiving strategy can be tailored to the level of end-user control and processing complexity associated with the spreadsheet. The following two categories are recommended:

1. Spreadsheets that do not, or only minimally, process data and which are created and maintained by end users, and

2. Spreadsheets that do more complex processing of data in order to perform or support critical processes and in which the IT department may have more involvement in the design and management.

Some examples based on the reconciliation of trading transactions will serve to illustrate what types of spreadsheets would fall into these two categories. Spreadsheets not used to process data include spreadsheets in which an individual compares reports from two systems and records any breaks as the list of exceptions for a particular day. Also included in this category would be a spreadsheet in which information is downloaded from a system and in which a pivot table is then used to create summary data. On the other had, spreadsheets that are used to process data would include a scenario in which information is downloaded from two systems into two separate MS Excel spreadsheets. MS Excel functions and pivot tables are then used to create summary data for each source. The data is then manually reconciled.

It is recommended that the archiving of spreadsheets that fall into the first category be dealt with in the same manner as the archiving of other unstructured content (e.g., Word files, some MS Access databases). Spreadsheets that perform more complex processing functions, on the other hand, are better handled as mini applications in which their archiving is dealt with in the context of managing the application life cycle.

## 10. ARCHIVING SPREADSHEETS AS UNSTRUCTURED CONTENT

Although, spreadsheets that fall into the first category of spreadsheet do not perform complex functions, they still do form an important part of the trail of evidence that SOX requires. For this reason, it is risky simply to rely on archiving the source data and recreating the spreadsheet in the event of a request for documentation, as investigators will be looking for evidence with integrity and authenticity (i.e., documentation produced contemporaneously in the normal and ordinary course of business). As such, it is a much less risky strategy to preserve, and be in a position to present, the entire evidentiary trail - source data and spreadsheet.

Generally speaking the creation and management of spreadsheets, that fall in the first category falls to the end user. To ensure that an organisation does not develop the problems experienced by the failed Jamaican banks, an organisation should establish some controls over how end users create and store spreadsheets. The following offers some examples of scaleable archiving strategies that an organization can pursue to implement spreadsheet archiving controls for spreadsheets that do not perform complex data processing:

1.  Low criticality/small scale operations - designate folder on server as archival folder and place all spreadsheets in folder in P1317 or PDFIA format to lock down content. Naming of files in the folder should be standardised, and documented controls should be established over who does the archiving, who has access to the folder, who can delete, etc. Ensure no deletions of files that fall within business critical/SOX relevant categories before their required retention period has been fully met or of any files if relevant to an ongoing or reasonably anticipated investigation, etc. Keep all files in online storage until retention period is expired, or if using removable storage media such as tape, establish a formalised programme to regularly review its integrity and refresh the medium or migrate content as necessary. Note: virus check before you put anything into your archival store to protect your archival records.

2.  High criticality/large scale operations. Introduce an electronic document management system with WORM storage. Institute Information Life Cycle Management (ILM) processes. The proper operation of EDRMS depends upon having a well-thought out and constructed corporate taxonomy, as it is the taxonomy which identifies the categories of business records that that the organisation creates and received, and the retention requirements that apply to each type. Organisations may want to look at taxonomy management software to support this, as taxonomy development and management can consume a large amount of resource. When an EDRM system is used to support spreadsheet archiving, end users (or the system if using an auto-classification feature) will associate the spreadsheet to an appropriate category in the taxonomy thereby ensuring that the spreadsheet will be retained for the period of time indicated by its association with a particular category.

3.  . Med. Criticality/med scale operations. Mix of 1 and 3

## 11. ARCHIVING SPREADSHEETS AS SOFTWARE ASSETS

Spreadsheets that fall into the second category of spreadsheet (i.e., those that perform more complex processing) may or may not be developed and/or supported by an IT department, depending on the organisational context. Regardless, given the function they perform, these spreadsheets may be treated more suitably as mini applications and their archiving dealt with in the context of managing the software life cycle.

Paying attention to data archiving requirements at the time at which a spreadsheet is created can make the process of archiving much easier and more effective in the long run. This is much easier to do if one applies a system development life cycle approach to the development of complex spreadsheets. It can be very useful to insert a records retention checkpoint at the system development project initiation phase. For example,

Spreadsheet designers could be asked whether they have identified the retention requirements for the spreadsheet and to outline how those retention requirements will be met. This will alert spreadsheet designers to the necessity of considering retention requirements and help them plan for data archiving.

Spreadsheet designers will be supported in their efforts if they can refer to organisation wide Records Retention Schedules that identify the retention requirements for given types of data and if the organisation has established standard data archiving solutions.

## 12. MIGRATING SPREADSHEETS OUT OF A PRODUCTION ENVIRONMENT TO AN ARCHIVE

It was once the case that organisations relied on backup tapes for both disaster recovery and retention purposes. It is now generally agreed that reliance on backup processes is no longer a suitable data archiving strategy. The problem with a reliance on backup processes, according to a Robert Frances Group research note is that administrators spend up to six hours per week recovering old messages for users, and responding to legal discovery can cost hundreds of thousands of pounds (RFG, 2005). Since the need to catalogue, locate and retrieve information in a timely manner has become more urgent in large part due to regulatory pressure, organisations have begun to embrace information life cycle management (ILM) in order to be able to free space for mission-critical data and provide an index and audit trail of archived information to support corporate governance.

This has given risen to a need for archiving tools that perform a long-term information and preservation access function; are able to keep up with steady input streams as well as that that have period inputs; allow a wide variety of organisational arrangements (i.e., local implementations covering inputs from one production system or a central archives covering inputs from several production systems). In response, software vendors have begun to offer archiving tools that can be used to handle the archiving of more complex and critical spreadsheets. These archiving tools are designed to take data from a production environment and migrate it to off-line storage for retention until that data is no longer needed. Compliance oriented archiving tools supply comprehensive storage, storage management, and security offerings to address data retention needs. In addition, a number of third-party archiving providers have emerged, so an organisation need not maintain its own archive.

When archiving spreadsheets to off-line and less expensive storage, the spreadsheet can be retained in its production format (e.g., an MS Excel file) or it can be converted to and retained in an open standard format such as PDF/A or WL to protect against technological change. The decision about the best format in which to retain the file should be in proportion to the expected length of time for which the spreadsheet must be retained, that is, the longer the retention, the better it will be to retain in an open format.

In terms of managing the migration of spreadsheet data from a production environment to a data archive environment, ISO 14721, the Open Archival Information System Standard, presents a reference model for the preservation of data that provides very useful guidance. It should be noted that the migration of data into archival systems needs to be tightly controlled in order to ensure that data is not lost and that data authenticity and integrity is maintained. For the same reason, control must be maintained over the management of the archival repository at all times. Here again, ISO 14721 is very instructive on this point.

## 12. CONCLUSION

Most discussions of spreadsheet risk focus on the factors contributing to accuracy and reliability of spreadsheet data content. Archiving, however, is often overlooked. But, as argued in this paper, it is critical for full SOX compliance. While SOX does not provide detailed guidance on corporate record keeping, the absence of such guidance should not be taken to mean that an organisation's leadership would not be expected to assess whether they have potential vulnerabilities and liabilities as a result of poor spreadsheet archiving practices and to take step to mitigate these. In their publication on the records management of Sarbanes-Oxley, authors John Montana, J. Edwin Dietal and Christine Martins write: ·All corporate recordkeeping is going to be under much closer and intense scrutiny in the future... One must be able to show that an aggressive, thoughtful, innovative corporate records and information management program is in place and continually being improved to ensure that individuals that might allege failure to comply with Sarbanes-Oxley are not successful." (Montana, et al., 2003)

APPENDIX A
RECORDS-RELATED REQUIREMENTS IN THE SARBANES-OXLEY ACT

4.  SOX 102(e) - Registration applications and annual reports must be available for public inspection subject to rules of the Board or Commission and applicable confidentiality laws.

5.  SOX 103(a)(2)(A)(i) - The Board shall establish quality control and ethical standards for registered public accounting firms in the preparation and issuance of audit reports; Audit reports, work papers, and other information related to any audit report must be kept for at least 7 years; Audit reports must contain statements about the testing of internal controls and whether those control structures include maintenance of accounting records.

6.  SOX 104(e) - Public accounting firms may be required to retain records not otherwise required under section 103.
    SOX 105(b)(2)(B)(C) & (D) - The Board may require production of audit work papers or any other documents in the possession of a registered public accounting firm or any other person, including any client of the firm.

8.  SOX 105(b)(5)(A) - All documents and information prepared by or given to the Board, related to an investigation under section 104, including Board deliberations, are confidential. (With certain enumerated exceptions under paragraph (B).)

9.  SOX 105 (c)(1) - The Board must keep a record of its proceedings.

10.  SOX 105(c )(5)(A) & (B) - Applies sanctions to both intentional and negligent conduct.

11.  SOX 106 - Foreign accounting firms that issue opinions or otherwise perform material services for a US company must supply audit work papers to the Board or Commission in connection with any investigation and be subject to the jurisdiction of US courts.

12.  SOX 201 - Amends section 1 OA of the Securities Exchange Act of 193 4 t prohibit registered public accounting firms from providing bookkeeping or other services related to accounting records or financial statements contemporaneously with audit services. It also precludes them from designing or implementing financial information systems at the same time, as well as performing other enumerated services contemporaneous with an audit.

13.  SOX 202 - Amends section I OA of the Securities Exchange Act of 1934 to require Audit Committees to preapproval all audit and nonaudit services with certain enumerated exceptions.

14.  SOX 204 - Amends section 1 OA of the Securities Exchange Act of 193 4 to require accounting firms to report to Audit Committees all critical accounting policies and practices to be sued, all alternative treatments of financial information that have been discussed and other "material written communications" between the accounting firm and the management of the company.

15. SOX 3 01 - Amends section 1 OA of the Securities Exchange Act of 1934 to make Audit Committees establish procedures for the receipt, retention, and treatment of complaints regarding accounting, internal accounting controls, or auditing matters. Does not list a period of years for retention. Audit Committees to establish retention conditions.

16. SOX 306 - Prohibits any director or executive officer from purchasing, selling, or otherwise acquiring or transferring nay equity security of the issuer during a blackout period if he/she acquires it in connection with his/her service or employment as a director or executive officer. Any profit realized by him/her will be recoverable by the issuer. Action to recover profits must be brought within 2 years of the date on which the profit was realized.

17. SOX 404 - Management Assessment of Internal Controls - many spreadsheets, will form a key component of being able to substantiate a company's financial statements and must therefore be available to the regulators if a company's statements are questioned. As noted by the U.S. Securities and Exchange Commission, 1ncreased retention of identified records also may provide critical evidence of financial reporting impropriety or deficiencies in the audit process." (SEC, 2003). Verification of a company's financial statements for a given period may take place a number of years after release, so a company should be prepared to produce supporting documents until certain the verification process is complete. There are serious penalties for making a false declaration - up to 10 years - so company directors and senior managers will want to be sure they can substantiate their financials.

18. SOX s. 802, rule 2-06(a) requires a 7 years after conclusion of the audit/review retention period for accountants to retain "records relevant to the audit or review of issuers' and registered investment companies' financial statements, including workpapers and other documents that form the basis of the audit or review, and memoranda, correspondence, communications, other documents and records (including electronic records), which are created, sent or received in connection with the audit or review." A company, in consultation with its legal and accounting advisors, will need to determine which of its spreadsheets and other records fall within the meaning of this provision of the act and related rules. There is a 10 year penalty for violating this rule.

19. SOX 802a - criminal penalties and their implications. "Whoever knowingly alters, destroys, mutilates, conceals, or makes a false entry in any record, document or tangible object with the intent to impede, obstruct, or influence the investigation or proper administration of any matter within the jurisdiction of any department or agency of the United States or any case filed under title 11, or in relation to or contemplation of any such matter or case, shall be fined under this title, imprisoned not more than 20 years, or both." This establishes the need for a legal hold regime so as to halt the deletion of any spreadsheet or other document that may be needed in case of litigation or investigation, even if only at the point of being anticipated.

20. SOX 906 - Requires written statements from CE0s and CF0s in annual reports to certify that the information is in compliance with the Securities Exchange Act of 1934 and that the report fairly represents the financial condition and results of operations of the issuer. "Criminal penalties of not more than $5,000,000, or

imprisonment of not more than 20 years, or both for wilfully certifying any statement, knowing that it does not comport with all the requirements set forth."

21. SOX 1102 - Which says that "Whoever corruptly (1) alters, destroys, mutilates, or conceals a record, document, or other object, or attempts to do so, with the intent to impair an object's integrity or availability for use in an official proceeding; or (2) otherwise obstructs, influences, or impedes any official proceeding, or attempts to do so, shall be fined under this title, or imprisoned for not more than 20 years, or both.

22. SOX 1106 - Amends section 32(a) of the Securities Exchange Act of 193 4 to increase the penalties of section 78ff(a)

23. SOX 1107 - Amends section 1513 of the title 18 United States Code to punish retaliation against informants by fines or imprisonment of not more than 10 years, or both. The amount of possible fines is not stated in the Act.